# Domain wall displacement in Py square ring for single nanometric magnetic bead detection

P. Vavassori[1], V. Metlushko[2], M. Gobbi[3], M. Donolato[3], M. Cantoni[3], R. Bertacco[3]

[1]CIC nanoGUNE Consolider, E-20009 San Sebastian, Spain and CNR-INFM National Research Center S3, CNISM and Dipartimento di Fisica, Università di Ferrara, I-44100 Ferrara, Italy

[2]Department of Electrical and Computer Engineering, University of Illinois at Chicago, Chicago, 60607 IL, USA

3 LNESS – Dipartimento di Fisica Politecnico di Milano, Via Anzani 42, 22100 Como (Italy)

## Abstract

A new approach based on the domain wall displacement in confined ferromagnetic nanostructures for attracting and sensing a single nanometric magnetic particles is presented. We modeled and experimentally demonstrated the viability of the approach using an anisotropic magnetoresistance device made by a micron-size square ring of Permalloy designed for application in magnetic storage. This detection concept can be suitable to biomolecular recognition, and in particular to single molecule detection.

Intense research is presently directed toward the development of high sensitivity magnetoresistive sensors for the detection of magnetic beads because of their potential high impact on biochemical applications and diagnosis in medicine. Magnetic beads can be prepared in sizes ranging from many micrometers to a few nanometers and they are primarily used as contrast agents in magnetic resonance imaging (MRI) and magnetic cell sorting. Furthermore, since the pioneering work by Baselt et al.[1] magnetic beads have been used for labeling and detecting target molecules at the surface of magnetic sensors where probe molecules are immobilized. The combination of functionalized magnetic beads and transducers then allows for the realization of magnetic biosensors for molecular recognition which hold promise to become key elements in biomedical research as well as in health care, pharmaceutical industry and environmental analysis applications. Due to the high sensitivity of spintronic transducers, stability of magnetic particles and absence of a significant magnetic background in most biological samples, high biological sensitivity can be achieved in immunoassays, theoretically down to the single molecule detection if nanobeads are employed. This is particularly appealing for biomedical research, offering the opportunity of immobilizing and detecting a biomolecule at a desired location for fundamental studies of its functions and dynamics.[2]

A number of very sensitive magnetic field detection devices have been developed during recent years able to detect a single micromagnetic particle, such as giant magnetoresistance (GMR) sensors,[1] spin-valves,[3] miniaturized silicon Hall sensors,[4] planar Hall effect sensors based on Permalloy thin films[5] and tunnelling magnetoresistance sensors (MTJs)[6]. The use of the anisotropic magnetoresistance (AMR) effect in ring-shaped sensors as sensitive bead detectors was suggested by Miller et al.[7] and, more recently the approach has been extended by J. Liandro et al.[8] to ring-shaped multilayered (pseudo-spin-valve) sensors based on GMR. In some cases the possibility of detecting single nanometric beads has been shown.[9] However, to date, none of these approaches has experimentally demonstrated recognition of single nanometric magnetic beads with diameter of the order of 100 nm.

Here we demonstrate the capability of detection of individual nanometric magnetic beads with a diameter of below 100 nm previously attracted to the active sensing area of a AMR sensor based on a ferromagnetic micron-size square ring. The reduced dimension of beads is of primary importance in biological assays oriented to single molecule detection for two main reasons. Firstly, it allows for the reduction of the perturbation on the affinity between the target and probe molecules. Furthermore if the tag dimensions are comparable to those of target biomolecules to be assayed, true single biomulecule recognition becomes accessible (limit of one molecule - one bead). Another salient feature of the sensor concept presented here is that the entire device does not need to be as small as the nano-beads to be sensed, making its fabrication less challenging. In fact the nanometric size of the sensing area arises from the physical properties of a domain wall (DW) localized at a geometric corner. The concept illustrated in the present paper relies on a previous experimental work made on square rings of Permalloy (Py) for application in magnetic storage of information.[10] In that design a tranverse head-to-head domain wall can be positioned at a given corner and its position can be read electrically thanks to the anisotropic magnetoresistance effect: when a DW is present between two sensing leads a reduction of the resistance is observed since some of the magnetization of the DW points perpendicularly to the current flow. Otherwise, when there is no DW present between the two sensing leads, the magnetization follows the direction of the perimeter of the ring and the resistance is higher. In this work we adapted this device to demonstrate a new detection concept suitable for the detection of magnetic nanobeads. The panel (a) of Fig. 1 shows the scanning electron microscopy image of the structure used in the present experiment. The 30 nm thick Py square ring structures have been lithographically patterned on top of 20 nm thick and 100 nm wide Au contacts, previously fabricated on a $SiO_2$/Si substrate. The outside size of the rings is 1.0 μm × 1.0 μm, the width of each segment is about 180 nm and the slit is about 80 nm wide. For the magnetoresistance measurements presented here, the voltage drop was measured using a lock-in amplifier between contacts labeled 3 and 4 in the panel (a) of Fig. 1, with a current of 15 μA injected at contact 1 while contact 2 was grounded. The schematic 3D shown in panel (b) of Fig. 1, illustrates the sensing concept proposed here: when a bead is placed over a domain wall previously positioned at one corner of the ring structure and a magnetic field H is applied to displace the DW along one ring edge, a magnetic dipole moment **μ** is generated in the superparamagnetic bead, as shown in the Figure. The stray field generated by **μ** opposes and partially cancels the applied field below the bead causing an increase of the value of the field H required to displace the DW. In panel (c) of Fig. 1 we present the result of a simulation of the effect when a magnetic nano-bead (a commercial MICROMOD nanomag®-D with diameter 130 nm and magnetic moment μ ≈ 150 x $10^{-15}$ emu at 1000 Oe) is placed over the DW at a vertical distance of 15 nm from the surface of the Py structure. 3D micromagnetic simulations of the whole system were performed by using the Object Oriented Micromagnetic Framework (OOMMF).[11] The material parameters used for the micromagnetic simulations are those contained in the OOMMF program for Py while for the magnetic bead we used a value of 130 emu/cm$^3$ for the magnetization at saturation that corresponds to the nominal value of 150 x $10^{-15}$ emu for the magnetic moment of the bead at 1000 Oe; the size of the unit cell used in the computation was 10x10x15 nm$^3$. In detail, the graphs in Fig. 1c show the variation of the magnetization component parallel to the applied field, normalized to its saturation value, in the upper segment of the ring as a function of the external field used to displace the DW, without (solid line) and with (dashed line) a bead over the structure. The simulations show that due to the proximity of the bead, the displacement of the DW is retarded by 12 Oe with respect to the case of the ring alone (compare dashed with solid line Fig. 1c) hereby demonstrating the potential for the utilization of this sensing concept for the detection of individual nanometric magnetic beads. A condition to be fulfilled for the correct operation of the device is that the nano-particle is placed exactly over one corner of the ring structure. Particles not sitting on top of the domain wall produce negligible changes in the displacement field. This is a problem common to all the magnetoresistive sensors oriented to single molecule detection developed so far for which an external action (magnetic fields, micromanipulators, …) is required to place and magnetize the particle for an optimal detection. Our approach is novel also in this respect since no external action is needed to place the bead over the corner, viz., the active area of the sensor, thanks to magnetic self focusing on the DW. This is illustrated in panels (a) and (b) of Fig. 2, which show atomic force microscopy (AFM) images taken from the ring structure after having first placed a DW in two opposite corners of the ring and subsequently dispensed the beads on the chip surface. The latter was capped with a 30 nm thick $SiO_2$ layer to avoid any local specific chemical interaction between beads and the different materials of the sensors. We typically dispensed a 1 µl drop of solution of beads in ethanol with final concentration of about $10^6$ particles/µl and then immediately dried the surface with nitrogen. Two types of commercial nano-beads MACS$^{TM}$ and nanomag®-D having diameter of 50 and 130 nm respectively were employed. The AFM image in panel (a) refers to the 50 nm beads and shows that a cluster of several beads is formed over one of the corner where a DW is located. The AFM image in panel (b) shows a single 130 nm bead placed over the opposite corner, where the other DW is located. We repeated the experiment several times changing either ring structure or the DW position and we always observed that clusters (most common for the 50 nm beads) or single particles are found over the corners where the DWs are located, but never over the other two corners of the ring.

This interesting behaviour can be understood by looking at the magnetic force microscopy (MFM) image in panel (c) of Fig. 2 showing that there is a magnetic field emanating from the structure only in correspondence of the DWs (left-top and bottom-right corners in the image). Elsewhere the field is negligible, apart from the region of the slit where a sizable field emanates in some rings, possibly due to the non planarity of the Py structure induced by Au contacts. The strong field gradient that characterizes this stray field causes a self-focusing action that can trap and drag towards the corner sensing region a nano-particle flowing in the vicinity of the structure. The force acting on a nano-particle at a certain distance from the ring surface can been calculated by computing with OOMMF the magnetic field **H** created in the surrounding space by the nano-structure in the magnetic configuration with a DW at the two opposite corners of the ring. Then the following vector expression is used for the force: $\mathbf{F} = -\mu_0(\boldsymbol{\mu}\cdot\boldsymbol{\nabla})\mathbf{H}$, where $\boldsymbol{\mu} = \mu(H)\mathbf{h}$ with $\mu(H)$ the known magnetization curve of the bead (provided by the manufacturer) and **h** is a unit vector parallel to the field **H**. Panel (d) of Fig. 2 shows the contour plot of the modulus of the force acting on a nano-bead of 130 nm-diameter computed in a plane at 200 nm from the sensor surface. The plot shows indeed that a force in range 0.1-10 pN is acting on the bead over an area of about 400 nm in diameter, assuring for an effective trapping and focusing action on the nano-bead.

Figure 3 shows the results of the magnetoresistance measurements carried out to verify the sensing concept presented here. Measurements were carried out onto rings without $SiO_2$ capping because with $SiO_2$ beads tend to move during the sweep of the magnetic fields, and reproducibility becomes a big challenge. Panel (a) shows the AFM image taken from a ring clean, while panel (b) shows the AFM image of the same ring after the dispensation of beads of 130 nm diameter (MICROMOD nanomag®-D) in ethanol and rinsing with deionised water for 1 minute. In panel (b) the white line indicates the position of the DW in the top-left corner of the ring structure, while the circles evidence that the original particles have broken up in four smaller beads, having an average diameter of about 80 nm, placed over and next to the corner. The fragmentation of the beads is not uncommon in these commercial particles that are indeed clusters of several smaller superparamagnetic nano-particles. For instance, a variation of the pH of the solution can easily cause this fragmentation. Furthermore fragmentation revealed to be relevant for the present experiment because a stronger chemical interaction between beads and the sensor surface took place. At variance with the case of intact beads on $SiO_2$, like that of figure 2b, the position of the fragments was stable during repeated magnetoresitive measurements thus allowing to prove the reproducibility of experiments. Note, in the meanwhile, that this is exactly what we would expect in case of a functionalised $SiO_2$ capping layer for biomolecular recognition due to the chemical affinity between target and probe molecules. Panel (c) of Fig. 3 shows the voltage drop measured across the corner (leads 2 and 3 in Fig. 1a) where the DW is initially positioned as a function of a magnetic field H applied parallel to the upper edge of the ring for the ring clean, as shown in panel (a), (square symbols) and with beads on top (circles) as shown in previous panel (b). The measurements were recorded by measuring a single loop starting from H = 0 (label A in the Figure), increasing H up to 250 Oe and then decreasing it down to -250 Oe and finally increasing H up to 0 Oe (final state marked by the label B in the Figure). The first transition observed in the voltage signal (resistance increase) when H is increased from 0 to 250 Oe corresponds to the displacement of the DW from the top- left to the top-right corner of the ring, while the second transition (resistance decrease) occurring when H is subsequently reduced to -250 Oe corresponds to the DW being brought back to the initial position. The magnetoresistance measurements show that, due to the proximity of the beads, the displacement of the DW from the top-left to the top-right corner is retarded by 12 Oe ($\Delta H$ in Fig. 3c). The fact that no variation of the DW displacement field is observed when the DW is brought back to the initial corner (transition for negative fields in figure 3) confirms that $\Delta H$ is due to the presence of the beads next to the initial corner and demonstrates the viability of the sensing concept proposed here. The chip has been finally rinsed in deionized water for 1 hour in order to completely remove the four fragments of beads from the corner, as checked by AFM. Magnetoresistance measurements gave exactly the same result obtained on the clean ring (curve with squares of figure 3), before dispensation of the beads, thus demonstrating the reliability of the observed change in the displacement field. Micromagnetic simulations of this geometry with four fragments of beads placed as in the experiment give a shift in the displacement field of the order of 12 Oe in good agreement with the experimental value This is obtained for the minimum beads-sensor distance compatible with the mesh (15 nm), while this value is unaltered when increasing the distance up to 30 nm.

Simulations also indicate that only the two beads over the DW have an influence, with the bead exactly on top of the DW (indicated by the arrow in figure 3) that is essentially responsible of the shift of 12 Oe, while the other one gives a contribution of only 2 Oe. In this sense our experiment demonstrate the detection of two beads, and also the possibility of single bead detection as the error in the evaluation of the displacement field is ± 2 Oe.

Moreover simulations show that the reduction of the width of the ring to the nano-bead diameter, which should also reduce the possibility of multiple bead clustering, doubles the value of ΔH. This in conjunction with the use of nano-beads of a higher magnetic moment (values up to 5 times that of the beads used here are reported in the literature for non-commercial nano-beads) can increase the achievable value of ΔH by about ten times with respect to the value obtained in this preliminary experiment. This is relevant in view of the necessity of capping the sensor surface with an insulating protecting layer functionalized with probe molecules (typically 100 nm thick) which causes an unavoidable decrease of the perturbation of the magnetic beads on the DW.

To summarize we demonstrated the viability of a novel approach for sensing single nanometric beads, based on their impact on the displacement field of a DW at a corner of a Py nanostructure. The self focusing effect on the active sensor area due to the strong magnetic field emanating from the DW has also been demonstrated. This proof of concept paves the way to the optimization of the sensor geometry for improving sensitivity and fulfilling the requirements imposed by biological recognition in view of application to single biomolecule detection and localization.

The authors thank P. Fraternali, S. Brivio and D. Petti for fruitful discussion and M. Leone for his skilful technical support. This work was supported by ... V.M. acknowledges support by the U.S. NSF, Grant ECCS-0823813 and by the U. S. Department of Energy, Office of Science, Office of Basic Energy Sciences, under Contract No. DE-AC02- 06CH11357 (CNM ANL grants Nr.468 and Nr.470).

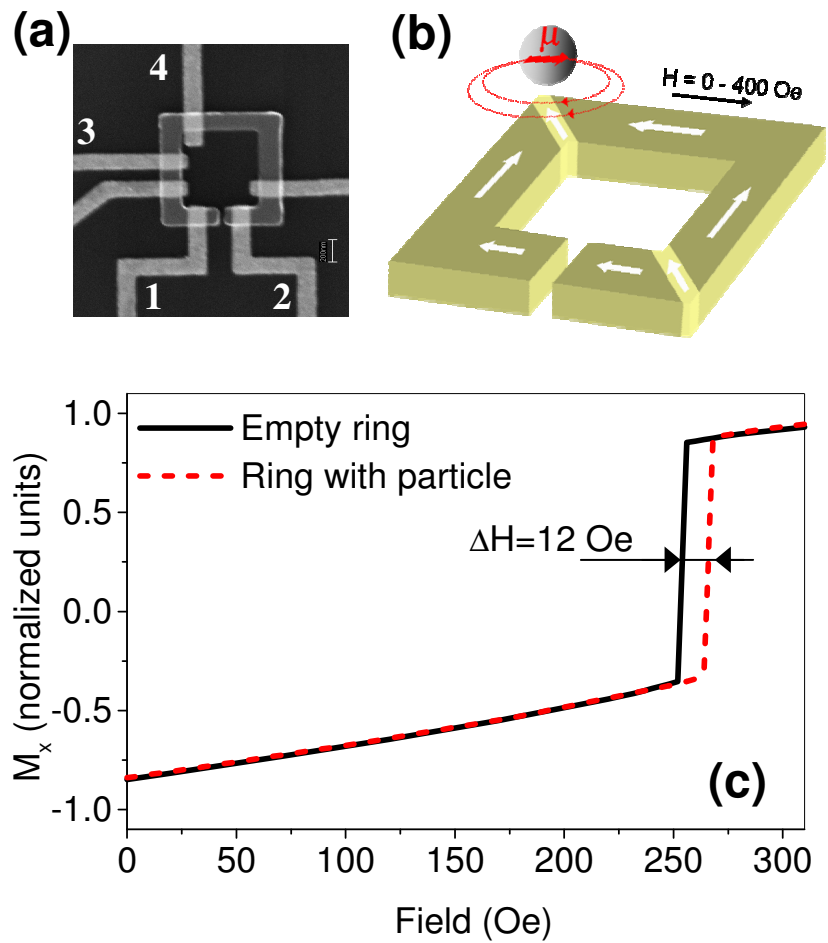

Figure 1

**Figure 1:** Panel (a): SEM image (2.5 μm x 2.5 μm) of the device structure. Panel (b): 3D schematic of the ring with two domain walls in the top-left and bottom-right corners and with a magnetized nanoparticle of moment μ producing a stray field in the top-left corner. Panel (c): domain wall displacement field calculated without (solid line) and with (dashed line) a magnetic particle of 130 nm placed at a distance of 15 nm over the corner with a domain wall. The magnetic field is applied as shown in panel (b).

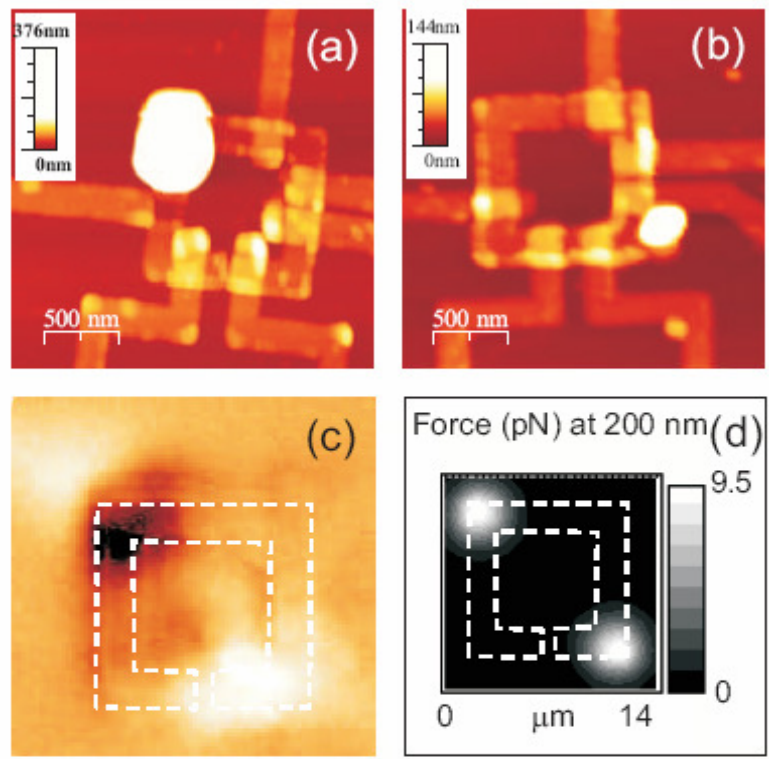

**Figure 2:** AFM images from rings capped with 30 nm of $SiO_2$ after beads have been dispensed in solution on the chip surface: (a) cluster of 50 nm beads, (b) single 130 nm bead. The diagonal lines indicates the position of the DW. Panel (c): MFM image from a ring after initialization with a DWs the top-left and bottom-right corners. Panel (d): simulation of the focusing magnetic forces on a plane at 200 nm distance from the ring surface. The dashed lines in panels (c) and (d) are guide for the eyes.

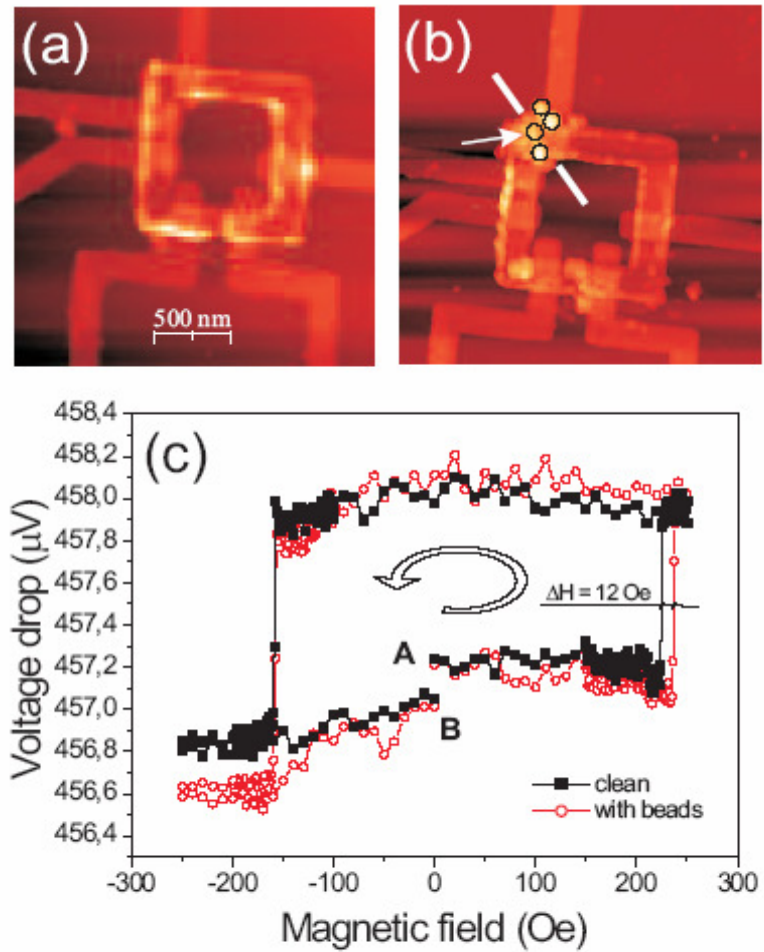

**Figure 3:** AFM images from a ring clean (a) and after beads dispensation (b). In panel (b) the white line indicate the position of the DW, while circles put in evidence four beads positioned in proximity of the corner, and the arrow indicates the bead substantially affecting the DW displacement field, according to micromagnetic simulations. Panel (c): voltage drop across the corner as a function of the magnetic field for the same ring clean (squares) and with beads on top (circles); labels A and B mark the beginning and the end of the acquisition, respectively.